\documentclass[superscriptaddress, preprint,a4paper]
{revtex4-1}
\usepackage{textcomp}
\usepackage{colordvi}
\usepackage{graphicx}
\usepackage{dcolumn}
\usepackage{bm} 
\usepackage{latexsym} 
\usepackage{longtable} 
\begin{document}
\pagestyle{myheadings}
\title{Evidence for a spin-aligned neutron-proton paired phase
from the level structure of $^{92}$Pd}
%
%
%
\newcommand{\nucl}[2]{\ensuremath{^{#1}}\mbox{#2}}

\author{B.~Cederwall\footnote[2]{Corresponding author: cederwall@nuclear.kth.se}}
\affiliation{\small
Department of Physics, Royal Institute of Technology, SE-10691 Stockholm, Sweden.}

\author{F.~Ghazi Moradi}
\affiliation{\small
Department of Physics, Royal Institute of Technology, SE-10691 Stockholm, Sweden.}

\author{T.~B\"ack}
\affiliation{\small
Department of Physics, Royal Institute of Technology, SE-10691 Stockholm, Sweden.}

\author{A.~Johnson} 
\affiliation{\small
Department of Physics, Royal Institute of Technology, SE-10691 Stockholm, Sweden.}

\author{J.~Blomqvist}
\affiliation{\small
Department of Physics, Royal Institute of Technology, SE-10691 Stockholm, Sweden.}

\author{E.~Cl\'ement} 
\affiliation{\small
Grand Acc\'el\'erateur National
d'Ions Lourds (GANIL), CEA/DSM - CNRS/IN2P3, Bd Henri Becquerel, BP
55027, F-14076 Caen Cedex 5, France.}

\author{G.~de~France} 
\affiliation{\small
Grand Acc\'el\'erateur National
d'Ions Lourds (GANIL), CEA/DSM - CNRS/IN2P3, Bd Henri Becquerel, BP
55027, F-14076 Caen Cedex 5, France.}

\author{R.~Wadsworth}
\affiliation{\small
Department of Physics, University of York, YO10 5DD York, UK.}

\author{K.~Andgren}
\affiliation{\small
Department of Physics, Royal Institute of Technology, SE-10691 Stockholm, Sweden.}

\author{K.~Lagergren}
\affiliation{\small
Department of Physics, Royal Institute of Technology, SE-10691 Stockholm, Sweden.}
\affiliation{\small
Joint Institute for Heavy-Ion Research, Holifield Radioactive Ion Beam Facility, Oak Ridge TN, 37831, USA.}

\author{A.~Dijon} 
\affiliation{\small
Grand Acc\'el\'erateur National
d'Ions Lourds (GANIL), CEA/DSM - CNRS/IN2P3, Bd Henri Becquerel, BP
55027, F-14076 Caen Cedex 5, France.}

\author{G.~Jaworski}
\affiliation{\small
Heavy Ion Laboratory, University of Warsaw, ul. Pasteura 5a, 02-093, Warsaw, Poland.}
\affiliation{\small
Faculty of Physics, Warsaw University of Technology, Koszykowa 75, 00-662, Warsaw, Poland.}

\author{R.~Liotta}
\affiliation{\small
Department of Physics, Royal Institute of Technology, SE-10691 Stockholm, Sweden.}

\author{C.~Qi}
\affiliation{\small
Department of Physics, Royal Institute of Technology, SE-10691 Stockholm, Sweden.}

\author{B.M.~ Nyak\'o}
\affiliation{\small
ATOMKI, H-4001 Debrecen, Hungary.}

\author{J.~Nyberg}
\affiliation{\small
Department of Physics and Astronomy, Uppsala University,
SE-75121 Uppsala, Sweden.}

\author{M.~Palacz}
\affiliation{\small
Heavy Ion Laboratory, University of Warsaw, ul. Pasteura 5a, 02-093, Warsaw, Poland.}

\author{H.~Al-Azri}
\affiliation{\small
Department of Physics, University of York, YO10 5DD York, UK.}

\author{G.~de~Angelis} 
\affiliation{\small
Instituto Nazionale di Fisica Nucleare, Laboratori Nazionali di Legnaro, I-35020 Legnaro, Italy.}

\author{A.~Ata\c{c}}
\affiliation{\small
Department of Physics, Ankara University, 06100 Tandogan Ankara, Turkey.}

\author{S.~Bhattacharyya\footnote[3]{Present address: V.E.C.C, 1/AF Bidhan Nagar, Kolkata 700064, India.}} 
\affiliation{\small
Grand Acc\'el\'erateur National
d'Ions Lourds (GANIL), CEA/DSM - CNRS/IN2P3, Bd Henri Becquerel, BP
55027, F-14076 Caen Cedex 5, France.}

\author{T.~Brock}
\affiliation{\small
Department of Physics, University of York, YO10 5DD York, UK.}

\author{J~.R.~Brown}
\affiliation{\small
Department of Physics, University of York, YO10 5DD York, UK.}

\author{P.~Davies}
\affiliation{\small
Department of Physics, University of York, YO10 5DD York, UK.}

\author{A.~Di Nitto}
\affiliation{\small
Dipartimento di Scienze Fisiche, Universit\`a di Napoli and Instituto Nazionale di Fisica Nucleare, I-80126 Napoli, Italy.}

\author{Zs.~Dombr\'adi}
\affiliation{\small
ATOMKI, H-4001 Debrecen, Hungary.}

\author{A.~Gadea}
\affiliation{\small
IFIC, CSIC, University of Valencia, Valencia, Spain.}

\author{J.~ G\'al}
\affiliation{\small
ATOMKI, H-4001 Debrecen, Hungary.}

\author{B.~Hadinia} 
\affiliation{\small
Department of Physics, Royal Institute of Technology, SE-10691 Stockholm, Sweden.}

\author{F.~Johnston-Theasby}
\affiliation{\small
Department of Physics, University of York, YO10 5DD York, UK.}

\author{P.~Joshi}
\affiliation{\small
Department of Physics, University of York, YO10 5DD York, UK.}

\author{K.~ Juh\'asz}
\affiliation{\small
Department of Information Technology, University of Debrecen, H-4010 Debrecen, Hungary.}

\author{R.~Julin}
\affiliation{\small
Department of Physics, University of Jyv{\"a}skyl{\"a}, FIN-40014 Jyv{\"a}skyl{\"a}, Finland.}

\author{A.~Jungclaus}
\affiliation{\small
Instituto de Estructura de la Materia, CSIC, E-28006 Madrid, Spain.}

\author{G.~ Kalinka}
\affiliation{\small
ATOMKI, H-4001 Debrecen, Hungary.}

\author{S.O.~Kara}
\affiliation{\small
Department of Physics, Ankara University, 06100 Tandogan Ankara, Turkey.}

\author{A. Khaplanov}
\affiliation{\small
Department of Physics, Royal Institute of Technology, SE-10691 Stockholm, Sweden.}

\author{J.~Kownacki}
\affiliation{\small
Heavy Ion Laboratory, University of Warsaw, ul. Pasteura 5a, 02-093, Warsaw, Poland.}

\author{G.~La~Rana}
\affiliation{\small
Dipartimento di Scienze Fisiche, Universit\`a di Napoli and Instituto Nazionale di Fisica Nucleare, I-80126 Napoli, Italy.}

\author{S.~M.~Lenzi}
\affiliation{\small
Diparimento di Fisica dell'Universit\`a di Padova and Instituto Nazionale di Fisica Nucleare, Sezione di Padova, I-35122 Padova, Italy.}

\author{J.~Moln\'ar}
\affiliation{\small
ATOMKI, H-4001 Debrecen, Hungary.}

\author{R.~Moro}
\affiliation{\small
Dipartimento di Scienze Fisiche, Universit\`a di Napoli and Instituto Nazionale di Fisica Nucleare, I-80126 Napoli, Italy.}

\author{D.~R.~Napoli}
\affiliation{\small
Instituto Nazionale di Fisica Nucleare, Laboratori Nazionali di Legnaro, I-35020 Legnaro, Italy.}

\author{B.~S.~Nara~Singh}
\affiliation{\small
Department of Physics, University of York, YO10 5DD York, UK.}

\author{A. Persson}
\affiliation{\small
Department of Physics, Royal Institute of Technology, SE-10691 Stockholm, Sweden.}

\author{F.~Recchia}
\affiliation{\small
Diparimento di Fisica dell'Universit\`a di Padova and Instituto Nazionale di Fisica Nucleare, Sezione di Padova, I-35122 Padova, Italy.}

\author{M.~Sandzelius\footnote[4]{Present address: Department of Physics, University of Jyv{\"a}skyl{\"a}, FIN-40014 Jyv{\"a}skyl{\"a}, Finland.}} 
\affiliation{\small
Department of Physics, Royal Institute of Technology, SE-10691 Stockholm, Sweden.}

\author{J.-N.~Scheurer} 
\affiliation{\small
Universit\'e Bordeaux 1, CNRS/IN2P3, Centre d'Etudes Nucl\'aires de Bordeaux Gradignan, UMR 5797, Chemin du Solarium, BP120, 33175 Gradignan, France.}

\author{G.~Sletten} 
\affiliation{\small
The Niels Bohr Institute, University of Copenhagen, 2100 Copenhagen, Denmark.}

\author{D.~Sohler}
\affiliation{\small
ATOMKI, H-4001 Debrecen, Hungary.}

\author{P.-A.~S{\"o}derstr{\"o}m}
\affiliation{\small
Department of Physics and Astronomy, Uppsala University,
SE-75121 Uppsala, Sweden.}

\author{M.~J.~Taylor}
\affiliation{\small
Department of Physics, University of York, YO10 5DD York, UK.}

\author{J.~Tim\'ar}
\affiliation{\small
ATOMKI, H-4001 Debrecen, Hungary.}

\author{J.~J.~Valiente-Dob\'on}
\affiliation{\small
Instituto Nazionale di Fisica Nucleare, Laboratori Nazionali di Legnaro, I-35020 Legnaro, Italy.}

\author{E.~Vardaci}
\affiliation{\small
Dipartimento di Scienze Fisiche, Universit\`a di Napoli and Instituto Nazionale di Fisica Nucleare, I-80126 Napoli, Italy.}

\author{S.~Williams}
\affiliation{\small
TRIUMF, Vancouver, BC, V6T 2A3, Canada.}
\maketitle
{\bf
 \small
The general phenomenon of shell structure in atomic nuclei has been understood
since the pioneering work of Goeppert-Mayer, Haxel, Jensen and Suess~\cite{Spin-orbit}.They
realized that the experimental evidence for nuclear magic numbers could be explained
by introducing a strong spin-orbit interaction in the nuclear shell model potential.
However, our detailed knowledge of nuclear forces and the mechanisms governing
the structure of nuclei, in particular far from stability, is still incomplete. In nuclei
with equal neutron and proton numbers ($N = Z$), the unique nature of the atomic
nucleus as an object composed of two distinct types of fermions can be expressed
as enhanced correlations arising between neutrons and protons occupying orbitals
with the same quantum numbers. Such correlations have been predicted to favor a
new type of nuclear superfluidity; isoscalar neutron-proton pairing~\cite{Eng66,Goo79,Eng97,Sat97,Civ97}, in addition
to normal isovector pairing (see Fig. 1). Despite many experimental efforts these
predictions have not been confirmed. Here, we report on the first observation of
excited states in $N = Z = 46$ nucleus $^{92}$Pd. Gamma rays emitted following the $^{58}$Ni($^{36}$Ar,2$n$)$^{92}$Pd fusion–evaporation reaction were identified using a combination of state-of-the-art high-resolution γ-ray, charged-particle and neutron detector systems. Our results reveal evidence for a spin-aligned, isoscalar neutron–proton coupling scheme, different from the previous prediction~\cite{Eng66,Goo79,Eng97,Sat97,Civ97}. We suggest that this coupling scheme replaces normal superfluidity (characterized by seniority coupling~\cite{Talmi,Talmi2}) in the ground and low-lying excited states of the heaviest N = Z nuclei. The strong isoscalar neutron-
proton correlations in these $N = Z$ nuclei are predicted to have a considerable impact
on their level structures, and to influence the dynamics of the stellar rapid proton
capture nucleosynthesis process.
}

\small 

For all known nuclei, including those residing along the $N=Z$ line up to around mass 80, a detailed analysis of their
properties such as binding energies~\cite{Ma00} and the
spectroscopy of the excited states~\cite{Af05} strongly
suggests that normal isovector ($T=1$) pairing is dominant at low excitation energies.
On the other hand, there are long standing predictions for a change in the heavier $N=Z$ nuclei from a
nuclear superfluid dominated by isovector pairing to a structure where isoscalar ($T=0$) neutron-proton (np) pairing has a
major influence as the mass number increases towards the exotic doubly magic nucleus
\nucl{100}{Sn}~\cite{Eng66,Goo79,Eng97,Sat97,Civ97}, the heaviest $N=Z$ nucleus to be bound.

$N=Z$ nuclei with mass number $> 90$ can only be produced in the laboratory
with very low cross sections and the associated problems of identifying the reaction products and their associated
$\gamma$ rays from the vast array of $N>Z$ nuclei that are present in much greater numbers from the reactions used
have prevented observation of their low-lying excited states until now.
In the present work the experimental difficulties have been overcome through the use of a
highly efficient, state-of-the-art detector system and a prolonged experimental running period.

Excited states in $^{92}$Pd were populated
following heavy-ion fusion-evaporation reactions at the
Grand Acc\'el\'erateur National d'Ions Lourds (GANIL), France. $^{36}$Ar ions, accelerated to a kinetic energy
of 111~MeV, were used to bombard an isotopically enriched (99.83\%) $^{58}$Ni target.
Light charged particles (mainly protons and alpha particles), neutrons  and $\gamma$ rays
emitted in the reactions were detected in coincidence. A schematic
layout of the experimental set-up is shown in Fig. 2.

The two-neutron (2n) evaporation reaction channel following formation of the
\nucl{94}{Pd} compound nucleus, leading to \nucl{92}{Pd},
was very weakly populated with a relative yield of less than $10^{-5}$ of the total fusion cross section.
Gamma rays from decays of excited states in \nucl{92}{Pd} were identified by comparing $\gamma$-ray
spectra in coincidence with two emitted neutrons and no charged particles with $\gamma$-ray
spectra in coincidence with other combinations of
neutrons and charged particles. The typical efficiency for detecting any charged particle
was 66\%. This rises to 88\% or higher if more than one such particle is emitted in the
population of a particular exit channel. The clean identification
of neutrons is crucial, since scattering of neutron events from one neutron detector
segment to another can be misinterpreted as two neutrons, thereby giving rise to background
from reaction channels where only one neutron has been emitted in $\gamma$-ray spectra
gated by two neutrons. But, since neutrons have a finite velocity the difference in
detection time is typically smaller for interactions resulting from two separate neutrons
compared to a single scattered neutron. Background contributions from neutron scattering
in 2n-gated spectra were significantly reduced by applying a criterion on the time difference in
the time of flight parameter, relative to the distance between the neutron detectors firing.
After such corrections the efficiency for correctly identifying both neutrons from a 2n-event was 3$\%$. 
Figure~\ref{fig:spectra} shows projected spectra from the charged particle-vetoed,
$2n$-selected $E_{\gamma}-E_{\gamma}$ coincidence matrix when $\gamma$ rays coincident
with the 874~keV, 912~keV and 750~keV transitions ( a), b), and c), respectively) assigned to  $^{92}$Pd are selected. 
By comparing spectra with and without the charged particle veto condition applied it is clear
that these $\gamma$ rays are not associated with emission of charged particles from the compound nucleus.
Fig.~\ref{fig:spectra}~d) shows a plot of the intensity ratios of the  874~keV, 912~keV and 750~keV $\gamma$ rays
in coincidence with two neutrons and one neutron, respectively,
proving that the $\gamma$ rays assigned to \nucl{92}{Pd} belong to the 2n-evaporation reaction channel.
An extensive literature search was also performed in order to exclude the
possibility that the $\gamma$ rays assigned to \nucl{92}{Pd} could be
due to the decay of excited states in some other nucleus. In particular, $\gamma$ rays from
reactions involving possible target impurities were taken into account. 
See Supplementary Information for further details on the data analysis.

The three most intense $\gamma$-ray transitions assigned to
$^{92}\mbox{Pd}$ (874~keV, 912~keV and 750~keV) have been ordered
into a ground state band based on their relative
intensities (caption, Fig \ref{fig:spectra}). Hence, the uncertainties
in the relative intensities of the $\gamma$-ray transitions translates  into a
corresponding uncertainty in their ordering, and consequently, also in the absolute position
of the $4^+$ state.
As shown in Fig \ref{fig:spectra}, these $\gamma$ rays form a
mutually coincident decay sequence. Although the
limited statistics precludes an accurate angular distribution analysis
and hence firm spin assignments it is likely that the
874~keV, 912~keV and 750~keV  $\gamma$-ray transitions constitute
a cascade of stretched E2 transitions depopulating the first excited
$2^+$, $4^+$, and $6^+$ states, respectively (Fig \ref{fig:spectra}, right panel).

Nuclei immediately below $^{100}$Sn on the Segr\'e chart, with $Z, N < 50$,
may show special structural features since the active
neutrons and protons here can move in many identical orbits. Here, for
the heaviest $N \approx Z$ nuclei, state-of-the-art shell model calculations predict ground-state and low-lying yrast structures based on spin-aligned systems of neutron-proton pairs to appear, similar to a
scenario proposed by Danos and Gillet more than four decades ago [11]. The np-paired ground-state
configuration emanates from the strong attractive interaction between g9=2 neutrons and protons in
aligned angular momentum (J = 9) coupling and is hence different from the predictions of a BCS
type of isoscalar np pairing condensate in N ¼ Z nuclei [2-6].
The shell model calculations were
performed using empirical two-body matrix elements in the
$f_{5/2}$~$p_{3/2}$~$p_{1/2}$~$g_{9/2}$ model space, see
Supplementary Information for details. 
In Fig.~\ref{fig:theory} we show the results of our calculation as well as
the corresponding experimental data resulting from the present work in
\nucl{92}{Pd} and from Refs. \cite{pd94,pd96} in \nucl{94,96}{Pd}. 
The level structure of  the semi-magic ($N=50$) nucleus  \nucl{96}{Pd}, with
four proton holes relative to the $Z=50$ closed
shell core, exhibits the typical traits of a nucleus in the normal isovector pairing phase for
which the seniority coupling scheme dominates:
A transition from the ground state to the first excited $2^+$ state requires the breaking of
one $g_{9/2}$ proton-hole pair, and therefore the energy spacing between these
two levels is rather large. The distance between the subsequent levels
gradually decreases as the angular momentum vectors of the $g_{9/2}$ quasiproton holes
align until the $8^+$ state is reached.
Here, the angular momentum vectors of one pair of proton holes are maximally aligned,
and in order to reach higher-lying states the other pair has to be
broken. This spin sequence terminates in the $12^+$ state, where all four
proton holes in the $g_{9/2}$ orbital are fully aligned. In order to
reach higher spin states, excitations must be made across the $N,Z=50$
shell gap, requiring much more energy.
In contrast, the calculated spectrum  of \nucl{92}{Pd}, with  four
proton holes and four neutron holes relative to the \nucl{100}{Sn}
core, show a nearly constant energy spacing between consecutive levels.
To examine the role played by the components of the np interaction
in this spectrum we performed the same calculation while including
only the $T=0$ component of the interaction matrix elements,
including only the $T=1$ components, or excluding all np interactions (i. e.
all $T_z$=0 components). As seen in Fig.~\ref{fig:theory}, the calculated spectrum of \nucl{92}{Pd}
for the latter case is very much like the spectrum of the closed neutron shell nucleus \nucl{96}{Pd}.
For the full calculation (case SM in Fig.~\ref{fig:theory}) the calculated spectra agree very well
with the experimental ones.
It is evident that the $T=0$ component of the np interactions plays a dominating role for the spectrum of
\nucl{92}{Pd} while such interactions between the valence
nucleons are absent in \nucl{96}{Pd}. The calculated wave functions for the ground state and low-
lying yrast states in 92Pd are completely dominated by the isoscalar np pairs in the spin-aligned
$J^{\pi} = 9^+$ coupling. The nucleus \nucl{94}{Pd} represents an interesting intermediate case. 

A simple semiclassical picture can help to illustrate why the yrast states in \nucl{92}{Pd} are nearly
equidistant, in terms of the overlap between the valence particle orbits. It is a consequence of
the variation of the angle between the angular momentum vectors of np hole pairs circling in one
direction and the angular momentum vectors of those circling in the opposite direction, as a function of the total angular momentum. This variation is approximately linear for small values of
angular momentum. This mechanism for generating the total angular momentum in the nucleus
is quite different from those present in normal superfluid nuclei. The regularly-spaced level sequence observed in the full calculation for \nucl{92}{Pd} is therefore a distinct signature of the spin-aligned
isoscalar mode, in the absence of collective vibrational excitations (see Supplementary Information
for further details). The fact that the ordering of the experimentally observed $\gamma$-ray transitions
is affected by some uncertainty does not change the interpretation of the data. The effect of a
different ordering would be a maximal change in the $2^+$ and $4^+$ energies by 124 keV and 162 keV,
respectively, in similar agreement with the theoretical prediction. The special topology of the
ground-state wave function predicted for \nucl{92}{Pd} is illustrated schematically in Fig. 4 ( top left), and
compared with the case for the normal pairing phase in \nucl{96}{Pd} (top right). In the spin-aligned np
paired phase the main component of the nuclear ground-state wave function can, in a semiclassical picture, be regarded as built of a system of deuteron-like np hole pairs spinning around the core,
each with maximum angular momentum. The special character of the wave function also implies
a deformed intrinsic structure.

While the experimental data presented in this Letter strongly suggest that a spin-aligned
neutron-proton paired phase is present in $^{92}$Pd, further experimental information is needed to confirm this interpretation. In particular measurements of particle transfer reactions, $B$(E2; $0^+\rightarrow2^+$)
values using Coulomb excitation, and precise mass measurements, would help to elucidate the
structural evolution of nuclei along the $N = Z$ line and to develop a better understanding of
neutron-proton correlations and their implications for nuclear shell structure far from stability.
This is also of importance for understanding reaction rates as well as the end point of the astro-
physical rp-process [14, 15], which have impact on the composition and X-ray burst profiles of
accreting neutron stars, and the nucleosynthesis of neutron deficient isotopes.

\vskip 0.3cm
{\bf Acknowledgements}
\vskip 0.2cm
This work was supported by the Swedish Research Council, the G\"oran Gustafsson Foundation,
the European Union 6th Framework Programme ``Integrating Infrastructure Initiative -
Transnational Access'', No.~506065~(EURONS), the Hungarian Scientific Research Fund, OTKA, under 
Contract Nos. K72566 and K68801, the UK Engineering and Physical Sciences Research Programme ``Interacting Infrastructure Initiative - Transnational Access'', the UK STFC, the Polish Ministry of Science and Higher Education
grant No. N N202 073935, the Spanish Ministerio de Ciencia e Innovaci\'on under Contract No. FPA2007-66069, and Ankara University BIYEP project No. DPT 2005120140.
The authors would like to thank the European Gamma-ray Spectroscopy Pool for use of the neutron
detector system, L. Einarsson and R. Sepp{\"a}l{\"a} for providing some of the targets used in this experiment and the GANIL
staff for technical support and for providing the $^{36}$Ar beam.

\bibliographystyle{unsrt}

\begin{thebibliography} {99}

\bibitem{Spin-orbit}
Goeppert Mayer, M. On Closed Shells in Nuclei. II. Phys. Rev. {\bf 75}, 1969-1970 (1949).

\bibitem{Goo79}
Goodman, A. L. Restoration of axial symmetry of the equilibrium shape of 24Mg by pairing correlations. Adv. Nucl. Phys. {\bf 11}, 260-263 (1979).
 
\bibitem{Eng66}
Engel, J., Langanke, K. \& Vogel, P. Pairing and Isospin Symmetry in Proton-Rich Nuclei. Phys. Lett. B{\bf 389}, 211-216 (1966). 

\bibitem{Eng97}
Engel, J. {\em et~al.}. Neutron-proton correlations in an exactly solvable model. Phys. Rev. C{\bf 55}, 1781-1788 (1997).

\bibitem{Sat97}
Satula, W. \& Wyss, R. Competition between T = 0 and T = 1 pairing in proton-rich nuclei. Phys. Lett. B{\bf 393}, 1-6 (1997). 

\bibitem{Civ97}
Civitarese, O., Reboiro, M. \& Vogel, P. Neutron-proton pairing in the BCS approach. Phys. Rev. C{\bf 56}, 1840-1843 (1997).

\bibitem{Talmi}
Talmi, I. Generalized seniority and structure of semi-magic nuclei. Nucl. Phys. A{\bf 172}, 1-24 (1971). 
\bibitem{Talmi2}
Generalized seniority states with definite isospin. Nucl. Phys. A{\bf 686}, 217-240 (2001).

\bibitem{Ma00}
Macchiavelli, A.O. {\em et~al.}. Is there np pairing in N=Z nuclei? Phys. Rev. C{\bf 61}, 041303(R) (2000).

\bibitem{Af05}
Afanasjev, A. \& Frauendorf, S. Description of rotating N=Z nuclei in terms of isovector pairing. Phys. Rev. C{\bf 71}, 064318 (2005).

\bibitem{Danos}
Danos, M. and Gillet, V. Stretch scheme, a shell model description of deformed nuclei. Phys. Rev. 161, 1034–1044 (1967)

\bibitem{pd94}
Marginean, N. \emph{et al.}. Yrast isomers in $^{95}$Ag, $^{95}$Pd,  and $^{94}$Pd.
Phys Rev. C{\bf 67}, 061301 (2003)


\bibitem{pd96}
Alber, D., Bertschat, H. H.,Grawe, H., Haas. H. \& Spellmeyer, B.
Nuclear structure studies of the neutron deficientN=50 nucleus $^{96}$Pd.
Z. Phys. A{\bf 332}, 129-135 (1989).

\bibitem{Sch01}
Schatz, H. {\em et~al.}. End Point of the rp Process on Accreting Neutron Stars.
Phys. Rev. Lett.{\bf 86}, 3471-3474 (2001).

\bibitem{Cle04}
Clement, R. R. C. et al. New approach for measuring properties of rp-process nuclei. Phys. Rev. Lett. 92, 172502 (2004)

\bibitem{Sch97}
Scheurer, J. N. et al. Improvements in the in-beam $\gamma$-ray spectroscopy provided by an ancillary detector coupled to a Ge -spectrometer: the DIAMANT-EUROGAM II example. Nucl. Instrum. Methods Phys. Res. A 385, 501–510 (1997)

\bibitem{Gal04}
G\'al, J. et al. The VXI electronics of the DIAMANT particle detector array. Nucl. Instrum. Methods Phys. Res. A 516, 502–510 (2004)

\bibitem{Ske99}
Skeppstedt, \"O. et al. The EUROBALL neutron wall design and performance tests of neutron detectors. Nucl. Instrum. Methods Phys. Res. A 421, 531–541 (1999)

\bibitem{Aza99}
Azaiez, F. EXOGAM: A γ-ray spectrometer for radioactive beams. Nucl. Phys. A 654, 1003c–1008c (1999)

\bibitem{Sim00}
Simpson, J. et al. The EXOGAM array: a radioactive beam gamma-ray spectrometer. Heavy Ion Phys. 11, 159–188 (2000); see also http://pro.ganil-spiral2.eu/laboratory/detectors/exogam

\bibitem{v46_1}
O'Leary, C.D. {\em et~al.}. Neutron-proton pairing, Coulomb effects and shape coexistence in odd-odd N=Z $^{46}$V. Phys. Lett. B{\bf 459}, 73-80 (1999).

\bibitem{v46_2}
Lenzi, S.M. {\em et~al.}. Band termination in the N=Z odd-odd nuclleus $^{46}$V
Phys. Rev. C {\bf 60}, 021303(R) (1999)

\bibitem{ru91}
Heese, J. {\em et~al.}. High spin states and shell model description of the neutron deficient nuclei $^{90}$Ru and $^{91}$Ru. Phys. Rev. C{\bf 49}, 1896-1903 (1994).

\bibitem{rh91}
Marginean, N. {\em et~al.}. Identification of excited states and shell model description of the N=Z+1
nucleus $^{91}$Rh. Phys. Rev. C{\bf 72}, 014302 (2005).







\end{thebibliography}

\clearpage
\begin{figure}[htbp]
\resizebox{\linewidth}{!}{\includegraphics{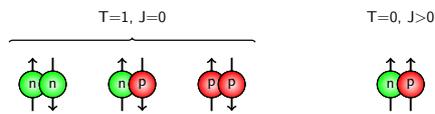}}
 \caption{Schematic illustration of the two possible pairing schemes in nuclei. The left part shows the normal isospin $T=1$
triplet . The two like-particle pairing components are responsible for most known effects of nuclear superfluidity.
Within a given shell these isovector components are restricted to spin zero due to the Pauli Principle.
The right part illustrates isoscalar $T=0$ neutron-proton pairing. Here the Pauli Principle allows only non-zero components
of angular momentum.
}
 \label{fig:np}
\end{figure}

\clearpage
\begin{figure}[htbp]
\resizebox{\linewidth}{!}{\includegraphics{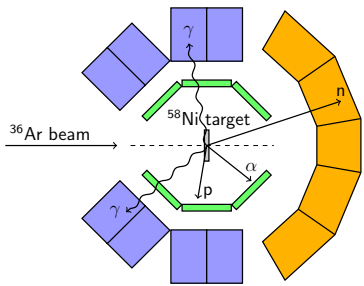}}
 \caption{Schematic illustration of the experimental set-up used to identify $\gamma$-ray transitions from excited states in $^{92}$Pd. The light particles and $\gamma$-rays emitted from the $^{36}$Ar + $^{58}$Ni reaction were observed using three different detector systems. The innermost detector array, DIAMANT\cite{Sch97,Gal04} (green), which consisted of 80 CsI scintillators, was used to detect light charged particles, mainly $\alpha$-particles and protons, and acted as a veto detector in the selection of events with no charged particles emitted. The Neutron Wall\cite{Ske99} (orange), comprising 50 liquid scintillator detectors and covering a solid angle of 1π in the forward direction, was used for the detection of evaporated neutrons. It is able to discriminate between neutron and $\gamma$-ray interactions by means of a combined time-of-flight and pulse-shape analysis technique. Gamma-rays emitted from the reaction products were detected using the EXOGAM\cite{Aza99,Sim00} high-purity Ge detector system (blue). Seven segmented Ge clover detectors were placed at an angle of 90$^\circ$ and four detectors at an angle of 135$^\circ$ relative to the beam direction, leaving room for the Neutron Wall at forward angles.
}
 \label{fig:np}
\end{figure}
\clearpage

\begin{figure}[htp]
\vspace{-3cm}
\resizebox{0.7\linewidth}{!}{\includegraphics{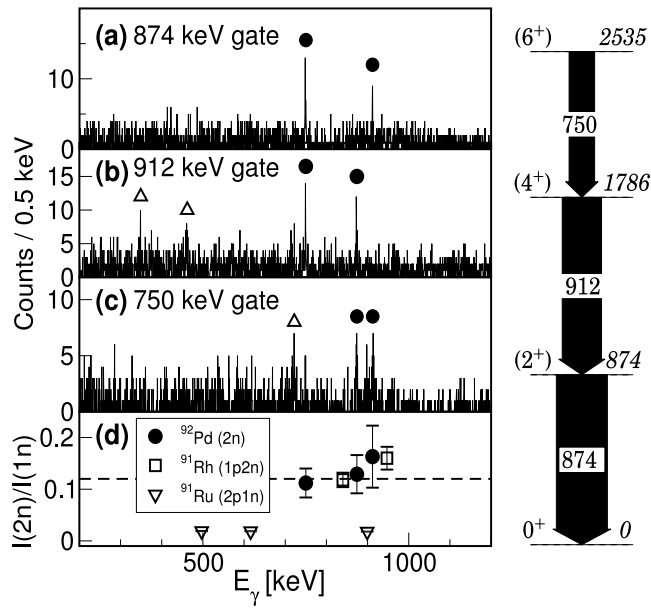}}
\vspace{-3cm}
 \caption{{\bf Left:}
{\bf (a), (b), and (c)} Gamma-ray energy spectra detected in coincidence with the
874~keV, 912~keV and 750~keV $\gamma$ rays, with the additional requirement that two neutrons and
no charged particle(s) were detected in coincidence. These $\gamma$ rays correspond to transitions
that we assign to depopulate the $2^+$,  $4^+$
and $6^+$ states in $^{92}$Pd, respectively and are marked by filled circles.
Gamma rays from \nucl{36}{Ar}-induced 1p1n-evaporation reactions on small amounts
of carbon deposited on the targets during irradiation, leading to
\nucl{46}{V} reactions  are visible in b) and c) (open triangles). These
$\gamma$ rays appear in the projected spectra due to a combined effect of the limited detection efficiency for charged particles,
the finite neutron/$\gamma$ separation in the neutron detectors, the presence of $\gamma$-ray transitions at 
914.9~keV and 750.7~keV in the level scheme of \nucl{46}{V}~\cite{v46_1, v46_2}, and the fact that the
reaction products from carbon contamination may recoil out of the target material, leading to Doppler broadening of
such $\gamma$ rays.
{\bf (d)} Intensity ratios of the $\gamma$ rays assigned to $^{92}$Pd in coincidence
with two neutrons and one neutron, respectively. The dashed line indicates the value expected for
$\gamma$ rays in coincidence with two neutrons, obtained from
the 1n- and 2n-detection efficiencies.
Measured intensity ratios for $\gamma$ rays from previously known
reaction products (\nucl{91}{Ru}~\cite{ru91} and \nucl{91}{Rh}~\cite{rh91}) are included for comparison.
{\bf Right:} Level scheme assigned to \nucl{92}{Pd}.
The energies (given in keV) and relative intensities (\%, normalized to the intensity of the 874~keV transition)
of the $\gamma$-ray transitions assigned to \nucl{92}{Pd} are as follows: 
873.6(2), 100(8);
912.4(2), 77(5);
749.8(3), 50(6). Given uncertainties are statistical.
}
\label{fig:spectra}
\end{figure}

\clearpage
\begin{figure}[htbp]
\resizebox{\linewidth}{!}{\includegraphics{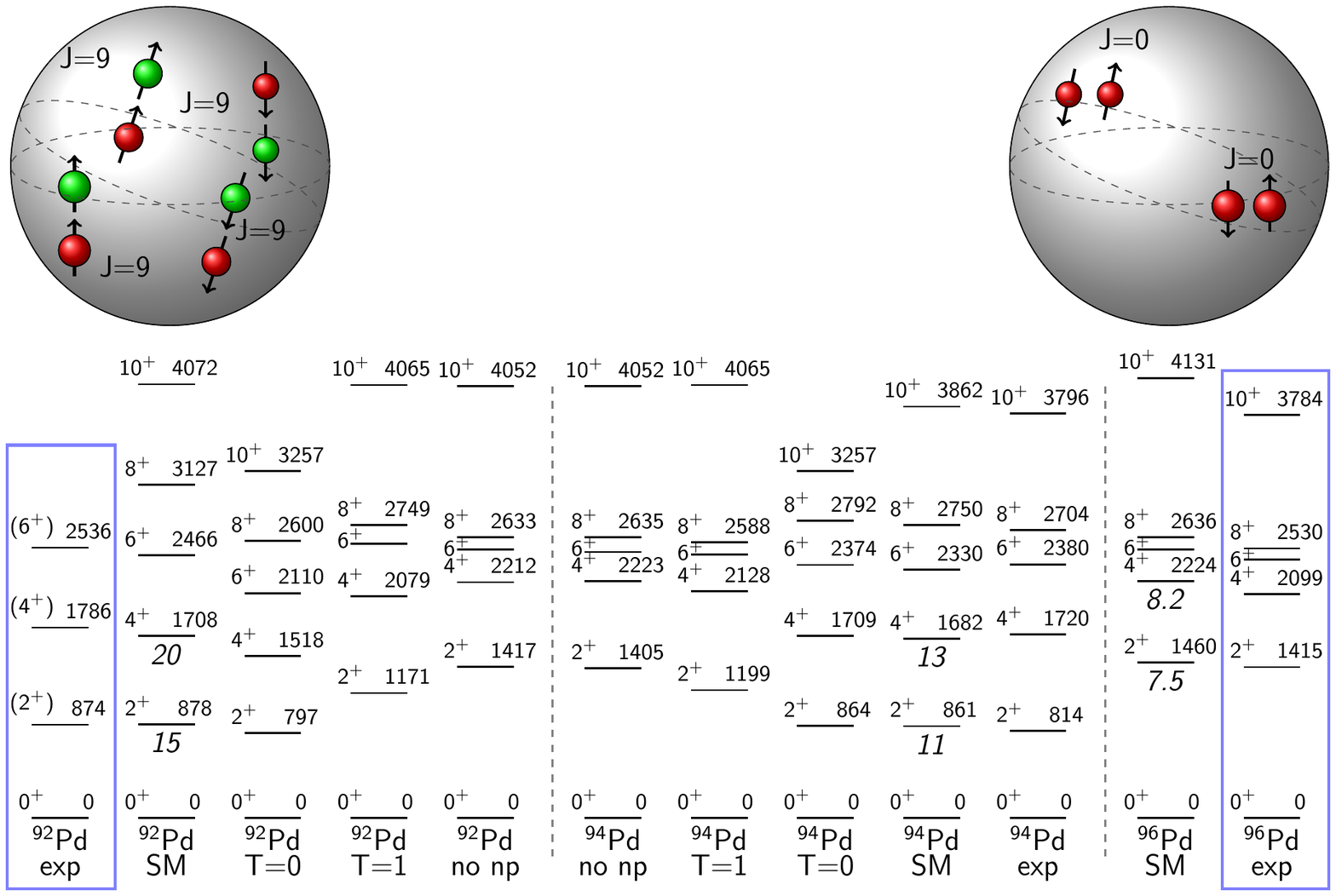}}
 \caption{{\bf Lower panel:} Experimental level energies (keV) in the ground state bands of \nucl{92}{Pd} (present work) 
and  \nucl{94,96}{Pd}~\cite{pd94,pd96}  compared with shell model predictions. Calculated
$B(E2: 2^+\rightarrow0^+)$  and $B(E2: 4^+\rightarrow2^+)$ values (W.u.) are also shown in italic below the
corresponding initial levels. The theoretical calculations for the spectra of
\nucl{92,94}{Pd} include, in addition to full neutron-proton
interactions (SM), also results for pure $T=0$ and pure $T=1$  neutron-proton interactions.
The results obtained without residual neutron-proton interactions (i.e. normal seniority coupling involving only
isovector, $T=1$, nn and pp pairing), are also shown for \nucl{92,94}{Pd}.\\
{\bf Top, left:} Schematic illustration of the structure of the ground-state wave function of \nucl{92}{Pd}
in the spin-aligned np paired phase. The main component of the wave function can be viewed as a system
of deuteron-like np hole pairs with respect to the $^{100}_{50}$Sn$_{50}$ ``core'', spinning around the centre
of the nucleus.
{\bf Top, right:} Schematic illustration of the structure of the ground-state wave function of \nucl{96}{Pd}
(normal pairing phase).
}
 \label{fig:theory}
\end{figure}

\end{document}